\documentclass[twocolumn,showpacs,floatfix,superscriptaddress,amsmath,amssymb,prl]{revtex4}
\usepackage{amsfonts}
\usepackage{bbm}
\usepackage{mathrsfs}
\usepackage{txfonts}
\usepackage{amssymb}
\usepackage{graphicx}
\usepackage{hyperref}
\usepackage{ulem}
 \usepackage{overpic}
 \usepackage{psfrag}
\usepackage{tabularx}
\usepackage{array}
\usepackage{placeins}
\makeatletter
\renewcommand\subsection{\@startsection
{subsection}{2}{0mm}
 {-\baselineskip}
 {0.5\baselineskip}
{\FloatBarrier\normalfont\Large\bfseries}}
\makeatother
\newcommand{\be}{\begin{equation}}
\newcommand{\ee}{\end{equation}}

\newcommand{\PreserveBackslash}[1]{\let\temp=\\#1\let\\=\temp}

\begin{document}
\title{Bifurcation in ground-state fidelity and universal order parameter for two-dimensional quantum transverse Ising model}

\author{Sheng-Hao Li} \affiliation{Centre for Modern Physics and Department of Physics,
Chongqing University, Chongqing 400044, The People's Republic of
China}

\author{Hong-Lei Wang} \affiliation{Centre for Modern Physics and Department of
Physics, Chongqing University, Chongqing 400044, The People's
Republic of China}

\author{Qian-Qian Shi} \affiliation{Centre for Modern Physics and Department of
Physics, Chongqing University, Chongqing 400044, The People's
Republic of China}

\author{Huan-Qiang Zhou}
\affiliation{Centre for Modern Physics and Department of Physics,
Chongqing University, Chongqing 400044, The People's Republic of
China}

\begin{abstract}
We establish an intriguing connection between quantum phase
transitions and bifurcations in the ground-state fidelity per
lattice site, and construct the universal order parameter for
quantum Ising model in a transverse magnetic field on an
infinite-size square lattice in two spatial dimensions, a
prototypical model with symmetry breaking order. This is achieved by
computing ground-state wave functions in the context of the tensor
network algorithm based on the infinite projected entangled-pair
state representation. Our finding is applicable to any systems with
symmetry breaking order, as a result of the fact that, in the
conventional Landau-Ginzburg-Wilson paradigm, a quantum system
undergoing a phase transition is characterized in terms of
spontaneous symmetry breaking captured by a local order parameter.
In addition, a bifurcation in the reduced fidelity between two
different reduced density matrices is also discussed.
\end{abstract}
 \maketitle
%%%%%%%%%%%%%%%%%%%%%%%%%%%%%%%%%%%%%%%%%%%%%
\section{I. Introduction}

In recent years, a novel approach to quantum phase transitions
(PQTs)~\cite{sachdev,wen} in quantum many-body lattice systems
emerges, based on the fact that fidelity, a basic notion in quantum
information science, is a measure of quantum state
distinguishability~\cite{zp,b6,zhou,zhou1,b7,fidelity1,fidelity2,b3,whl,rams}.
As argued in Refs.~\cite{b6,zhou,zhou1,b7,b3,whl}, the ground-state
fidelity per lattice site captures drastic changes of the
ground-state wave functions around a critical point for a quantum
system in any spatial dimensions. That is, it is a universal marker
to detect QPTs, regardless of what type of the internal order is
present.  In fact, quantum many-body lattice systems with either
spontaneous symmetry breaking (SSB) order or topological order, have
been specifically studied: for the former, the examples include
quantum transverse Ising~\cite{b3} and Potts chains~\cite{dai},
quantum Ising~\cite{b7} and XYX models~\cite{libo} on a square
lattice, while for the latter, the Kitaev model~\cite{b8} on a
honeycomb lattice. Remarkably, there is a smoking-gun signature for
SSB order in the fidelity per site approach: SSB order implies a
bifurcation, arising from degenerate ground-state wave functions due
to broken symmetry, in the ground-state fidelity per lattice site.

In fact, it was shown that, for a QPT arising from an SSB, a
bifurcation appears in the ground-state fidelity per site, with a
critical point identified as a bifurcation point. This in turn
results in a novel concept of the universal order parameter, which
appears as the ground-state fidelity per lattice site between a
ground state and its symmetry-transformed counterpart. The advantage
of the ground-state fidelity per lattice site and the universal
order parameter~\cite{c1} over local order parameters lies in the
fact that both of them are universal, in the sense that it is not
model-dependent, in contrast to model-dependant order parameters in
characterizing QPTs in quantum lattice many-body systems. Similarly,
a bifurcation occurs in the ground-state reduced
fidelity~\cite{Zhoufidelity} between the one-site reduced density
matrices and the two-site reduced density matrices, with a critical
point identified  as a bifurcation point. However, all specific
examples, up to now, have been restricted to quantum many-body
lattice systems in one spatial dimension.

In this paper, we take one step further to see if these novel ideas
are practical to quantum many-body lattice systems in two spatial
dimensions. We establish, by exploiting the tensor network algorithm
based on the infinite projected entangled-pair states (iPEPS)
algorithm~\cite{Jordan}, an intriguing connection between a QPT and
a bifurcation in the ground-state fidelity per lattice site, and
construct the universal order parameter for quantum transverse Ising
model on a square lattice, a prototypical model with symmetry
breaking order. Our finding is applicable to any systems with
symmetry breaking order, as a result of the fact that, in the
conventional Landau-Ginzburg-Wilson paradigm, a quantum system
undergoing a phase transition is characterized in terms of SSB
captured by a local order parameter. In addition, a bifurcation in
the reduced fidelity between two different reduced density matrices
is also discussed.

\section{II.  Model}
We consider quantum Ising model in a transverse magnetic field on an
infinite-size square lattice in two spatial dimensions. It is
described by the Hamiltonian:
\begin{equation}\label{fitising}
H=- \sum_{\langle ij \rangle}\sigma_{x}^{i} \sigma_{x}^{j} + \lambda
\sum_{i}\sigma_{z}^{i},
\end{equation}
where $\sigma_{\alpha}^{i} \; (\alpha = x, z)$  are the spin $1/2$
Pauli operators at site $i$, $\langle ij \rangle$ runs over all the
possible nearest-neighbor pairs on a square lattice, and $\lambda$
is  the transverse magnetic field, which we choose as a control
parameter. The model is invariant under the symmetry operation:
$\sigma_{x}^{[i]} \rightarrow -\sigma_{x}^{[i]}$ and
$\sigma_{z}^{[i]} \rightarrow \sigma_{z}^{[i]}$ for all sites
simultaneously, which yields the $Z_{2}$ symmetry. As is well known,
the system undergoes a second-order QPTs at the critical field
$\lambda_c \sim 3.044$~\cite{Jordan,QMC}.

\section {III. Bifurcation in the ground-state fidelity per lattice
site} For two ground-state wave functions $|\Psi(\lambda_1)\rangle$
and $|\Psi(\lambda_2)\rangle$ corresponding to two different values
$\lambda_1$ and $\lambda_2$ of the control parameter $\lambda$, the
ground-state fidelity
$F(\lambda_1,\lambda_2)=|\langle\Psi(\lambda_2)|\Psi(\lambda_1)\rangle|$
asymptotically scales as $F(\lambda_1,\lambda_2)\sim
d(\lambda_1,\lambda_2)^N$, with $N$ being the number of sites in the
lattice. Here, $d(\lambda_1,\lambda_2)$ is the scaling parameter,
introduced for one-dimensional quantum systems~\cite{b6,zhou,zhou1}
and for two and higher-dimensional quantum systems~\cite{b7}. Note
that it characterizes how fast fidelity goes to zero when the
thermodynamic limit is approached. Physically, the scaling parameter
$d(\lambda_1,\lambda_2)$ is the averaged fidelity per lattice site,
which is  well defined in the thermodynamic limit:
\begin{equation}
\ln d(\lambda_1,\lambda_2)\equiv \lim_{N\rightarrow \infty}\frac{\ln
F(\lambda_1,\lambda_2)}{N}.
\end{equation}
As noted in Refs.~~\cite{b6,zhou,zhou1,b7}, it satisfies the
properties inherited from the fidelity $F(\lambda_1,\lambda_2)$: (i)
normalization $d(\lambda,\lambda)= 1$; (ii) symmetry
$d(\lambda_1,\lambda_2)= d(\lambda_2,\lambda_1)$; and (iii) range $0
\leq d(\lambda_1,\lambda_2) \leq 1$.

In the $Z_{2}$ symmetric phase, the ground-state wave function is
non-degenerate, while in the $Z_{2}$ symmetry-broken phase, two
degenerate ground-state wave functions arise. If we choose
$|\Psi(\lambda_{2}) \rangle$ as a reference state, with
$\lambda_{2}$ in the $Z_{2}$ symmetric phase, then the ground-state
fidelity per lattice site, $d(\lambda_{1}, \lambda_{2})$, is not
able to distinguish two degenerate ground-state wave functions in
the $Z_{2}$ symmetry-broken phase. However, if we choose
$|\Psi(\lambda_{2}) \rangle$ as a reference state, with
$\lambda_{2}$ in the $Z_{2}$ symmetry-broken phase, then the
ground-state fidelity per lattice site, $d(\lambda_{1},
\lambda_{2})$, can be used to distinguish two degenerate
ground-state wave functions.  Therefore, a bifurcation occurs in the
ground-state fidelity per lattice site, $d(\lambda_{1},
\lambda_{2})$, as a function of $\lambda_1$, for a fixed
$\lambda_2$. As argued in Refs.~\cite{b3,whl}, for a given
truncation dimension $\mathbb{D}$, a pseudo-phase-transition point
$\lambda_{\mathbb{D}}$ manifests itself as a bifurcation
point~\cite{bifurcation}.

%%%%%%%%%%%%%%%%%%%%%%%%%%%%%%%%%%%%%%%%%%%%%%%%%%%%%%%%%%%%%%%%%%%%%
\begin{figure}
\vspace*{4cm}
  \includegraphics{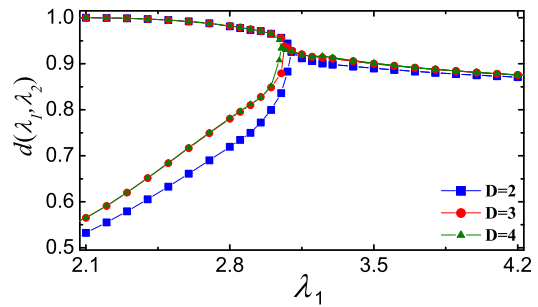}
\caption{(color online) The ground-state fidelity per lattice site,
$d(\lambda_{1},\lambda_{2})$, for quantum Ising model in a
transverse field on a square lattice in two spatial dimensions, with
the transverse magnetic field $\lambda$ as the control parameter. If
we choose $|\Psi(\lambda_{2}) \rangle$ as a reference state, with
$\lambda_{2}$ being in the $Z_{2}$ symmetry-broken phase, then
$d(\lambda_{1},\lambda_{2})$ distinguishes two degenerate
ground-state wave functions, with a pseudo-phase-transition point
$\lambda_{\mathbb{D}}$ as a bifurcation point. Here, we have chosen
$\lambda_{2}=2.1$. The pseudo-phase-transition point is identified
at $\lambda_{\mathbb{D}}=3.100$ for the truncation dimension
$\mathbb{D}=2$,  at $\lambda_{\mathbb{D}}=3.065$ for the truncation
dimension $\mathbb{D}=3$, and  at $\lambda_{\mathbb{D}}=3.050$ for
the truncation dimension $\mathbb{D}=4$, respectively.}
\label{fidelity}
\end{figure}
%%%%%%%%%%%%%%%%%%%%%%%%%%%%%%%%%%%%%%%%%%%%%%%%%%%%%%%%%%%%%%%%%%%%%

In Fig.~\ref{fidelity}, we plot the ground-state fidelity per
lattice site, $d(\lambda_{1},\lambda_{2})$, for quantum Ising model
in a transverse magnetic field on an infinite-size square lattice in
two spatial dimensions,  with the transverse magnetic field
$\lambda$ as the control parameter. Note that a bifurcation does
occur for the ground-state fidelity per lattice site,
$d(\lambda_{1}, \lambda_{2})$. Here, we have chosen
$\lambda_{2}=2.1$ as a specific example. The pseudo-phase-transition
point locates at $\lambda_{\mathbb{D}}=3.100$ for the truncation
dimension $\mathbb{D}=2$, at $\lambda_{\mathbb{D}}=3.065$ for the
truncation dimension $\mathbb{D}=3$, and  at
$\lambda_{\mathbb{D}}=3.050$ for the truncation dimension
$\mathbb{D}=4$. We mention that the ground-state fidelity per
lattice site, $d(\lambda_{1}, \lambda_{2})$, is computed from the
iPEPS representation of the ground-state wave functions, following
the transfer matrix approach described in Ref.\cite{b7}.

A remarkable feature of the bifurcation points for the ground-state
fidelity per lattice site, $d(\lambda_{1},\lambda_{2})$, as seen in
Fig.~\ref{fidelity}, is that  $d(\lambda_{1},\lambda_{2})$ between
different symmetry breaking ground-state wave functions in the same
phase appears less than that between two ground-state wave functions
from different phases. This is due to the fact that two degenerate
symmetry breaking ground states in the same symmetry-broken phase
are more distinguishable than two ground states from different
phases.

\section {IV. The ground-state reduced fidelity between two reduced density matrices}
For quantum Ising model in a transverse magnetic field on an
infinite-size  square lattice, the one-site reduced density matrix
in the $Z_{2}$ symmetric phase takes the form,
\begin{equation}\label {fitsonex}
   \rho = \frac{1}{2} + 2\langle S_{z}\rangle S_{z},
 \end{equation}
where $\langle S_{z}\rangle$ is the ground-state expectation value
of $S_{z}$ in the $Z_{2}$ symmetric phase, while the two-site
reduced density matrix is,
\begin{eqnarray}\label {fitwo}
   \rho &=& \frac{1}{4}I + 4\gamma_{xx} S_{x} \otimes S_{x} + 4\gamma_{zz} S_{z} \otimes S_{z} \notag\\
       && + \gamma_{oz} I \otimes S_{z} + \gamma_{zo} S_{z} \otimes
       I.
\end{eqnarray}
Here, $\gamma_{xx}=\langle S_{x} \otimes S_{x} \rangle$,
$\gamma_{zz}=\langle S_{z} \otimes S_{z} \rangle$,
$\gamma_{oz}=\langle I \otimes S_{z} \rangle$, $\gamma_{zo}=\langle
S_{z} \otimes I \rangle$, and $I$ is the identity matrix.

In the $Z_{2}$ symmetry-broken phase, the one-site reduced density
matrix becomes,
\begin{equation}\label {fitanone}
   \rho = \frac{1}{2} + 2\langle S_{x}\rangle S_{x} + 2\langle S_{z}\rangle
   S_{z},
 \end{equation}
whereas the two-site reduced density matrix is
\begin{eqnarray}\label {fitantwo}
   \rho &=& \frac{1}{4}I + 4\gamma_{xx} S_{x} \otimes S_{x} + 4\gamma_{zz} S_{z} \otimes S_{z} \notag\\
       && + \gamma_{oz} I \otimes S_{z} + \gamma_{zo} S_{z} \otimes I + 4\gamma_{xz} S_{x} \otimes S_{z} \notag\\
       && + 4\gamma_{zx} S_{z} \otimes S_{x} + \gamma_{ox} I \otimes S_{x} + \gamma_{xo} S_{x} \otimes
       I,
\end{eqnarray}
with $\gamma_{xz}=\langle S_{x} \otimes S_{z} \rangle$,
$\gamma_{zx}=\langle S_{z} \otimes S_{x} \rangle$,
$\gamma_{ox}=\langle I \otimes S_{x} \rangle$, and
$\gamma_{xo}=\langle S_{x} \otimes I \rangle$.

The reduced fidelity measures the distance between two quantum mixed
states. For two reduced density matrices $\rho_{\lambda_{1}}$ and
$\rho_{\lambda_{2}}$, the reduced fidelity
$F(\rho_{\lambda_{1}},\rho_{\lambda_{2}})$ is defined to
be~\cite{Zhoufidelity}
\begin{equation}\label {fit1}
   F(\rho_{\lambda_{1}},\rho_{\lambda_{2}})=tr\sqrt{ \rho_{\lambda_{1}}^{1/2}  \rho_{\lambda_{2}}
   \rho_{\lambda_{1}}^{1/2}}.
 \end{equation}
Here, $\rho_{\lambda_{1}}$ and $\rho_{\lambda_{2}}$ are the reduced
density matrices corresponding to two different values,
$\lambda_{1}$ and $\lambda_{2}$, of the control parameter $\lambda$.
Notice that the reduced fidelity
$F(\rho_{\lambda_{1}},\rho_{\lambda_{2}})$ is a function of
$\lambda_{1}$ and $\lambda_{2}$, which satisfies the following
properties: (i) normalization $F(\rho_{\lambda},\rho_{\lambda})=1$;
(ii) symmetry
$F(\rho_{\lambda_{1}},\rho_{\lambda_{2}})=F(\rho_{\lambda_{2}},\rho_{\lambda_{1}})$;
(iii) range 0$ \leq F(\rho_{\lambda_{1}},\rho_{\lambda_{2}})\leq $1.

In Fig.~\ref{one-site} (upper panel), we plot the ground-state
reduced fidelity $F(\rho_{\lambda_{1}},\rho_{\lambda_{2}})$ between
the one-site reduced density matrices for quantum Ising model in a
transverse field on an infinite-size square lattice, with the
transverse field strength $\lambda$ as the control parameter. Here,
we choose $\rho_{\lambda_{2}}$, at $\lambda_{2}=2.1$, as a reference
state, which breaks the $Z_{2}$ symmetry. The one-site reduced
fidelity is able to distinguish two mixed states (described by two
reduced density matrices) from two degenerate ground-state wave
functions, with a bifurcation point as a pseudo-phase-transition
point $\lambda_{\mathbb{D}}$. When the control parameter $\lambda$
crosses a pseudo-transition point, the ground-state degeneracy
changes suddenly, implying that the system undergoes a QPT. We
observe that the pseudo-phase-transition point
$\lambda_{\mathbb{D}}$ moves toward the critical point $\lambda_c$
with increasing $\mathbb{D}$. More precisely, the
pseudo-phase-transition point locates at
$\lambda_{\mathbb{D}}=3.100$ for the truncation dimension
$\mathbb{D}=2$,  at $\lambda_{\mathbb{D}}=3.065$ for the truncation
dimension $\mathbb{D}=3$, and at $\lambda_{\mathbb{D}}=3.050$ for
the truncation dimension $\mathbb{D}=4$, respectively.

We also plot the ground-state reduced fidelity
$F(\rho_{\lambda_{1}},\rho_{\lambda_{2}})$ between the two-site
reduced density matrices for quantum Ising model in a transverse
field on an infinite-size square lattice in
Fig.~\ref{one-site}\;(lower panel). The reference state is chosen at
the same $\lambda_{2}=2.1$, as in the case of the one-site reduced
fidelity. We observe that a bifurcation also occurs in the two-site
reduced fidelity, with the same pseudo-phase-transition point
$\lambda_{\mathbb{D}}$. This is expected, simply because they are
resulted from the same set of the ground-state wave functions.
%%%%%%%%%%%%%%%%%%%%%%%%%%%%%%%%%%%%%%%%%%%%%%%%%%%%%%%%%%%%%%%%%%%%%
\begin{figure}
\vspace*{4cm}
  \includegraphics{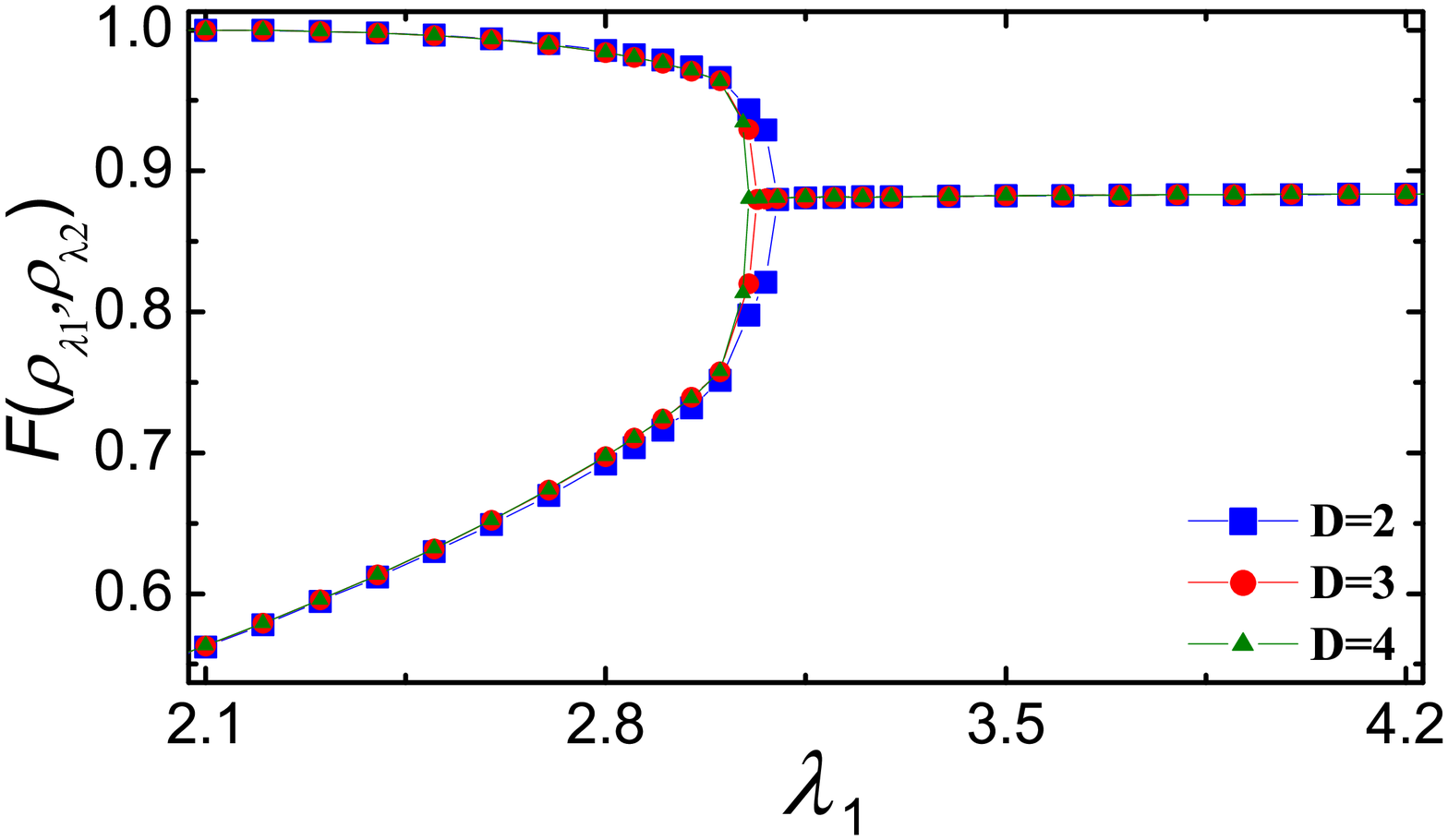}
\vspace*{4.2cm}
  \includegraphics{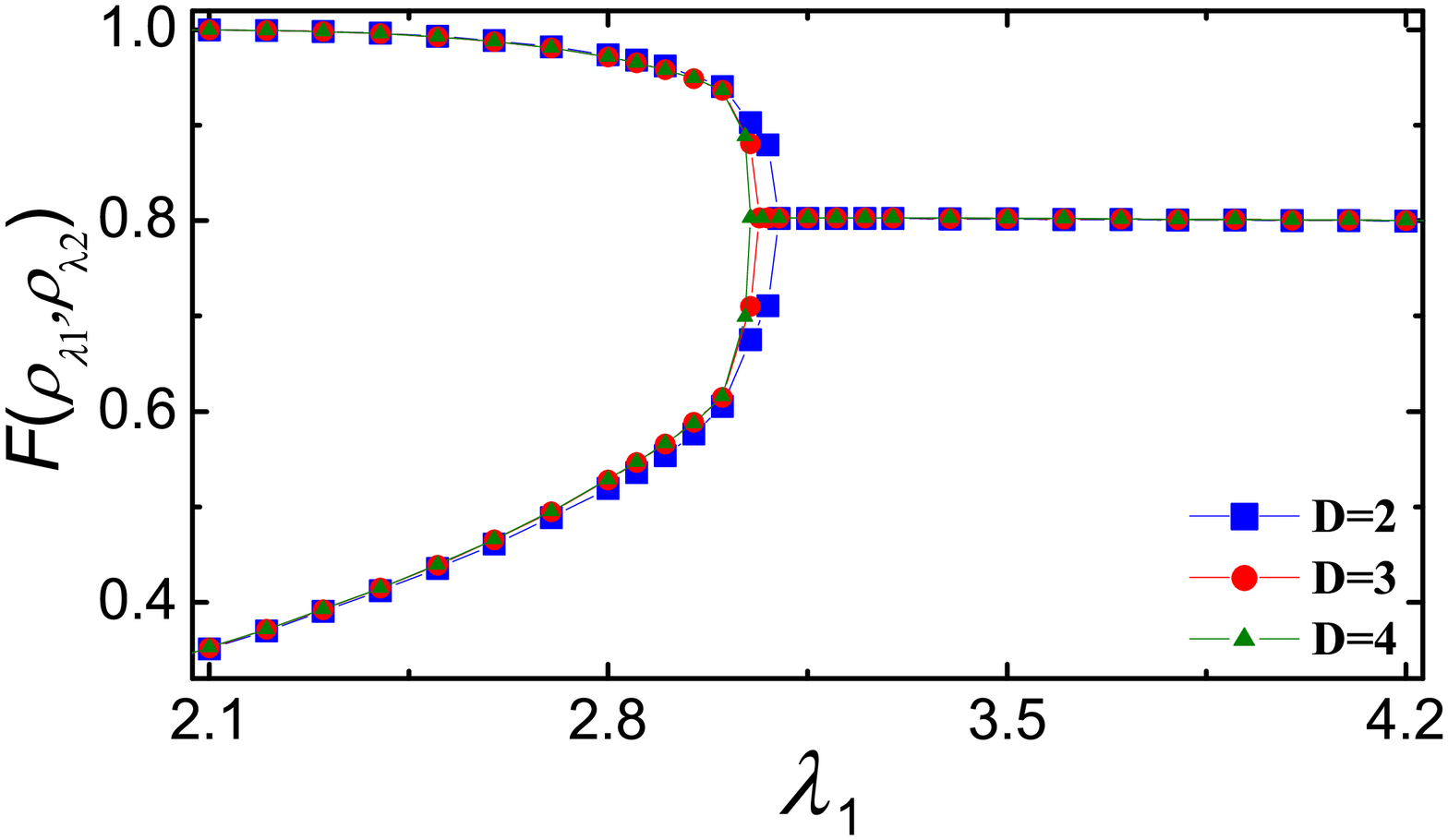}
\caption{(color online) Upper panel: The ground-state one-site
reduced fidelity, $F(\rho_{\lambda_{1}},\rho_{\lambda_{2}})$, for
quantum Ising model in a transverse field on a square lattice in two
spatial dimensions. Here, we have chosen $\rho_{\lambda_{2}}$ as the
reference state, with $\lambda_{2} = 2.1$ in the $Z_{2}$
symmetry-broken phase.   Lower panel: The ground-state two-site
reduced fidelity, $F(\rho_{\lambda_{1}},\rho_{\lambda_{2}})$, for
quantum Ising model in a transverse field on a square lattice. The
reference state $\rho_{\lambda_{2}}$  has been chosen at the same
$\lambda_{2} = 2.1$ as for the one-site reduced fidelity. Note that
$F(\rho_{\lambda_{1}},\rho_{\lambda_{2}})$ is able to distinguish
two mixed states arising from degenerate ground-state wave
functions. A pseudo-phase-transition point occurs as a bifurcation
point. In both cases, the pseudo-phase-transition point locates at
$\lambda_{\mathbb{D}}=3.100$ for the truncation dimension
$\mathbb{D}=2$, at $\lambda_{\mathbb{D}}=3.065$ for the truncation
dimension $\mathbb{D}=3$, and at $\lambda_{\mathbb{D}}=3.050$ for
the truncation dimension $\mathbb{D}=4$, respectively.}
\label{one-two}
\end{figure}
%%%%%%%%%%%%%%%%%%%%%%%%%%%%%%%%%%%%%%%%%%%%%%%%%%%%%%%%%%%%%%%%%%%%%

\section {V. The universal order parameter}
As argued in Ref.~\cite{c1}, for any quantum lattice system with a
symmetry group $G$ undergoing a QPT with symmetry breaking order,
there is a universal order parameter: it is defined as the ground
state fidelity per lattice site between a ground-state wave function
and its symmetry-transformed counterpart, which is discontinuous for
first-order phase transitions and continuous for second-order phase
transitions. This is based on the observation that, for any ground
state $|\Psi \rangle$ in the symmetric phase, $\langle \Psi |g |\Psi
\rangle$ is equal to 1, for any symmetry operation $g \in G$,
whereas it is identical to zero for any state in the symmetry broken
phase.

In order to measure the distance between two quantum states
$|\Psi(\lambda)\rangle$ and $g|\Psi(\lambda)\rangle$, let us
consider their counterparts $|\Psi(\lambda)\rangle_L$ and
$g|\Psi(\lambda)\rangle_L$ on a finite-size lattice, with $L$ being
the number of the total lattice sites. As argued in
Refs.~\cite{b6,zhou,zhou1,b7}, $|_L\langle \Psi |g |\Psi \rangle_L|$
asymptotically scales as $f_g^L(\lambda)$ with $L$, as one may see
from the tensor network representations of the system's ground state
wave functions. Here, $f_g(\lambda)$ is the averaged fidelity per
lattice site, which is well-defined even in the thermodynamic limit.
As such, one sees that, $f_g(\lambda)=1$ for any $g \in G$, if
$\lambda$ is in the symmetric phase $\lambda
>\lambda_c$, and $0<f_g(\lambda)<1$ for any nontrivial symmetry operation $g$, if $\lambda$ is in the
symmetry-broken phase $\lambda <\lambda_c$. As argued in
Ref.~\cite{c1}, we define the universal order parameter to be
\begin{equation}\label {fit2}
   I_g(\lambda)=\sqrt{1-f_g^2(\lambda)}.
 \end{equation}
Note that $I_g(\lambda)$ is always zero if $\lambda
>\lambda_c$. However, it becomes nonzero, with its value ranging from 0 to 1, if
$\lambda <\lambda_c$. These features are exactly what one requires
for $I_g(\lambda)$ to be an order parameter. In fact, this is valid
for any quantum many-body lattice system with a global symmetry
group $G$ spontaneously broken.  A remarkable feature of  the
universal order parameter is that it not only makes it possible to
locate a critical point, but also enables us to identify a
factorized state $|\Psi(\lambda_f) \rangle$, with $\lambda_f$ being
the so-called factorizing field~\cite{factorized}.

\begin{figure}
  \vspace*{4cm}
  \includegraphics{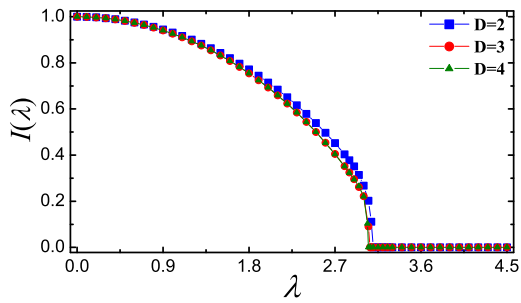}
\caption{(color online) The universal order parameter $I(\lambda)$
for quantum Ising model in a transverse magnetic field on a square
lattice.  A pseudo-phase-transition point $\lambda_{\mathbb{D}}$
occurs, as $I(\lambda)$ changes from being nonzero to zero at
$\lambda = \lambda_{\mathbb{D}}$. When the truncation dimension
$\mathbb{D}$ is increased, a pseudo-phase-transition point
$\lambda_{\mathbb{D}}$ approaches the critical point $\lambda_{c}$.
In addition, the universal order parameter $I(\lambda)$ reaches the
maximum value $I(\lambda)=1$ at the factorizing field
$\lambda_{f}=0$.} \label{super}
\end{figure}

In Fig.\ref{super}, we plot the universal order parameter
$I_g(\lambda)$ for quantum Ising model in a transverse field on an
infinite-size square lattice, with the field strength $\lambda$ as
the control parameter. Here, the symmetry operation is the
non-trivial element $g$ of the group $Z_2$. If $\lambda <
\lambda_{\mathbb{D}}$, the universal order parameter $I_g(\lambda)$
is non-zero. This characterizes the $Z_{2}$ symmetry-broken phase,
in contrast to the fact that the universal order parameter
$I_g(\lambda)$ is zero, in the symmetric phase $\lambda >
\lambda_{\mathbb{D}}$. When the control parameter $\lambda$ varies
across the pseudo-critical point $\lambda_{\mathbb{D}}$, the
behavior of the universal order parameter $I_g(\lambda)$ changes
qualitatively, implying that the system undergoes a QPT at the
pseudo-phase-transition point $\lambda_{\mathbb{D}}$. As the
truncation dimension ${\mathbb{D}}$ is increased, the
pseudo-phase-transition point $\lambda_{\mathbb{D}}$ moves toward
the critical point $\lambda_c$. In addition, the (trivial)
factorizing field $\lambda_f=0$ exists for quantum Ising model in a
transverse field on an infinite-size square lattice, at which
$I_g(\lambda)$ reaches its maximum.

%%%%%%%%%%%%%%%%%%%%%%%%%%%%%%%%%%%%%%%%%%%%%%%%%%%%%%%%%%%%%%%%%%%%%
%%%%%%%%%%%%%%%%%%%%%%%%%%%%%%%%%%%%%%%%%%%%%%%%%%%%%%%%%%%%%%%%%%%%%

\section {VI. Conclusions}

We have investigated an intriguing connection between QPTs and
bifurcations in the ground-state fidelity per lattice site, in the
context of the tensor network algorithm based on the iPEPS
representation. For quantum transverse Ising model on an
infinite-size square lattice, the iPEPS algorithm produces two
degenerate symmetry-breaking ground-state wave functions arising
from the $Z_2$ symmetry breaking, each of which results from a
randomly chosen initial state. Therefore, a quantum system
undergoing a phase transition is characterized in terms of SSB that
is captured by a bifurcation in the ground-state fidelity per
lattice site. We have also constructed the universal order parameter
, and discussed bifurcations in the ground-state reduced fidelity
between two different reduced density matrices in the
symmetry-broken phase, for quantum Ising model in a transverse
magnetic field on an infinite-size square lattice. We expect that
our approach might provide further insights into critical phenomena
in quantum many-body lattice systems in condensed matter.

\section {VII. Acknowledgments}
We thank Sam Young Cho, Bing-Quan Hu, Bo Li, Jin-Hua Liu, Yao-Heng
Su, and Jian-Hui Zhao for helpful discussions. This work is
supported in part by the National Natural Science Foundation of
China (Grant No: 10874252). SHL, HLW, and QQS are supported by the
Fundamental Research Funds for the Central Universities (Project No.
CDJXS11102214), and by Chongqing University Postgraduates' Science
and Innovation Fund (Project No.: 200911C1A0060322).


\begin{thebibliography}{10}

\bibitem{sachdev} S. Sachdev, Quantum Phase Transitions, Cambridge University
Press, 1999, Cambridge.
\bibitem{wen} X.-G. Wen, Quantum Field Theory of Many-Body Systems, Oxford
University Press, 2004, Oxford.

\bibitem{zp} P. Zanardi and N.
Paunkovi\'{c}, Phys. Rev. E  \textbf{74}, 031123 (2006).

\bibitem{b6} H.-Q. Zhou and J.P. Barjaktarevi\'c, J. Phys. A: Math. Theor. \textbf{41}, 412001 (2008).

\bibitem{zhou} H.-Q. Zhou, J.-H. Zhao, and B. Li, J. Phys. A: Math. Theor. \textbf{41}, 492002 (2008).

\bibitem{zhou1} H.-Q. Zhou, arXiv:0704.2945.

\bibitem{b7} H.-Q. Zhou, R. Or\'us, and G. Vidal, Phys. Rev. Lett. \textbf{100}, 080601 (2008).

\bibitem{b3} J.-H. Zhao, H.-L. Wang, B. Li, and H.-Q. Zhou, Phys. Rev. E \textbf{82}, 061127 (2010).

\bibitem{whl} H.-L. Wang, J.-H. Zhao, B. Li, and H.-Q. Zhou,
arXiv:0902.1670.

\bibitem{fidelity1} P. Zanardi, M. Cozzini, and P. Giorda, J. Stat. Mech. L02002, (2007);
N. Oelkers and J. Links, Phys. Rev. B \textbf{75}, 115119 (2007); M.
Cozzini, R. Ionicioiu, and P. Zanardi, Phys. Rev. B \textbf{76},
104420 (2007); L. Campos Venuti and P. Zanardi, Phys. Rev. Lett.
\textbf{99}, 095701 (2007); T. Liu, Y.-Y. Zhang, Q.-H. Chen, and
K.-L. Wang, Phys. Rev. A \textbf{80}, 023810 (2009).

\bibitem{fidelity2}
 W.-L. You, Y.-W. Li, and S.-J. Gu, Phys. Rev. E \textbf{76},
022101 (2007); S. J. Gu, H. M. Kwok, W. Q. Ning, and H. Q. Lin,
Phys. Rev. B \textbf{77}, 245109 (2008); M. F. Yang, Phys. Rev. B
\textbf{76}, 180403(R) (2007); Y. C. Tzeng and M. F. Yang, Phys.
Rev. A \textbf{77}, 012311 (2008); J. O. Fj{\ae}restad, J. Stat.
Mech.: Theory Exp. (2008) P07011; J. Sirker, Phys. Rev. Lett.
\textbf{105}, 117203 (2010).

\bibitem{rams} M.M. Rams and B. Damski, Phys. Rev. Lett. \textbf{106}, 055701 (2011).

\bibitem{dai} Y.-W. Dai, B.-Q. Hu, J.-H. Zhao and H.-Q. Zhou, J. Phys. A: Math. Theor. \textbf{43}, 372001 (2010).

\bibitem{libo} B. Li, S.-H. Li and H.-Q. Zhou, Phys. Rev. E \textbf{79}, 060101(R) (2009).

\bibitem{b8} J.-H. Zhao, and H.-Q. Zhou, Phys. Rev. B \textbf{80}, 014403 (2009).

\bibitem{c1} J.-H. Liu, Q.-Q. Shi, H.-L. Wang, and H.-Q. Zhou, arXiv:0909.3031.

\bibitem{Zhoufidelity}J.-H. Liu, Q.-Q. Shi, J.-H. Zhao, and H.-Q. Zhou, arXiv:0905.3031.

\bibitem{Jordan}J. Jordan, R. Or\'{u}s, G. Vidal, F. Verstraete, and J. I.
Cirac, Phys. Rev. Lett. \textbf{101}, 250602 (2008).

\bibitem{QMC} H. W. J. Blote and Y. Deng, Phys. Rev. E \textbf{66}, 066110 (2002).

\bibitem{bifurcation} J. D. Crawford, Rev. Mod. Phys. \textbf{63}, 991 (1991); J. Araki \textit{et. al.}, Proc.
R. Soc. Lond. A \textbf{345}, 413 (1975).

\bibitem{factorized} J. Kurmann, H. Thomas, and G. M\"uller, Physica A (Amsterdam) \textbf{112}, 235 (1982); C. Hoeger, G. von Gehlen, and V. Rittenberg, J. Phys. A:
Math. Gen. \textbf{18}, 1813 (1985);  V. Kendon, K. Nemoto, and W.
J. Munro, J. Mod. Opt. \textbf{49}, 1709 (2002); T. Roscilde, P.
Verrucchi, A. Fubini, S. Haas, and V. Tognetti, Phys. Rev. Lett.
\textbf{93}, 167203 (2004);  L. Amico, F. Baroni, A. Fubini, D.
Patan\'e, V. Tognetti, and P. Verrucchi, Phys. Rev. A \textbf{74},
022322 (2006); R. Oliveira, O. C. O. Dahlsten, and M. B. Plenio,
Phys. Rev. Lett. \textbf{98}, 130502 (2007); F. Baroni, A. Fubini,
V. Tognetti, and P. Verrucchi, J. Phys. A: Math. Theor. \textbf{40},
9845 (2007); S. Dusuel and J. Vidal, Phys. Rev. B \textbf{71},
224420 (2005); R. Rossignoli, N. Canosa, and J. M. Matera, Phys.
Rev. A \textbf{77}, 052322 (2008);  S. M. Giampaolo, G. Adesso, and
F. Illuminati, Phys. Rev. Lett. \textbf{100}, 197201 (2008).

\end{thebibliography}
\end{document}